\documentclass[journal abbreviation, manuscript]{copernicus}

\nolinenumbers

\begin{document}

\title{A Practical Introduction to Regression-based Causal Inference in Meteorology (II): Unmeasured confounders}


\Author[1,2][marzban@stat.washington.edu]{Caren}{Marzban} 
\Author[1]{Yikun}{Zhang}
\Author[3]{Nicholas}{Bond}
\Author[4]{Michael}{Richman}

\affil[1]{Department of Statistics, University of Washington, Seattle, Washington, 98195 USA}
\affil[2]{Applied Physics Laboratory, University of Washington, Seattle, Washington, 98195 USA}
\affil[3]{Climate Impacts Group, University of Washington, Seattle, Washington, 98195 USA}
\affil[4]{School of Meteorology, University of Oklahoma, Norman, Oklahoma, 73019 USA}



\runningtitle{Regression-based Causal Inference II}

\runningauthor{Marzban et al.}

\received{}
\pubdiscuss{} 
\revised{}
\accepted{}
\published{}


\firstpage{1}

\maketitle

\begin{abstract}
One obstacle to ``elevating'' correlation to causation is the phenomenon of confounding, i.e., when a correlation between two variables exists because both variables are in fact caused by a third variable, called a confounder. The situation where the confounders are measured is examined in an earlier, accompanying article. Here, it is shown that even when the confounding variables are not measured, under certain conditions it is still possible to estimate the causal effect via a regression-based method that uses the notion of instrumental variables. Using a meteorological data set, similar to that in the sister article, a number of different estimates of the causal effect are compared and contrasted. It is shown that the instrumental-variable estimates of causal effect depend on the choice of the instrumental variable, and that meteorological considerations are important in resolving the ambiguity. R code is provided for generating all of the results, and numerous directions for future work are outlined.
\end{abstract}

\vspace{1.0cm}

\centerline{\includegraphics[height=1.0in, width=1.5in]{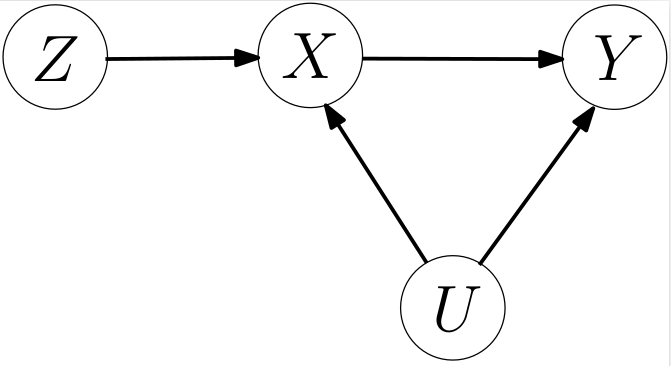}}


\introduction  

There exist several frameworks for causal inference; they are generally referred to as the potential
outcomes framework \citep{rubin1974}, the structural equations paradigm \citep{pearl2009}, and the 
graphical model approach \citep{pearl2010}. The pros and cons of these frameworks have been discussed 
thoroughly in textbooks on causal inference \citep{ding_book, hernan_robins, morgan_book, gelman_book}; 
and some work exists that aims to unify these approaches \citep{richardson_robins, richardson_etal}.
In the simplest situation, the goal is to assess the causal effect of a variable, called
{\it treatment}, on another variable, called {\it outcome}.

Although this article deals with the situation where the treatment is continuous, consider the
simpler case where it is binary (i.e., where there exist two possible values of the treatment).
Then, in an experimental setting, if/when the two treatment values can be randomly
assigned to the units on which the outcome variable is measured, causality can be readily assessed
by comparing the mean of the outcome variable across the two treatment groups; this set-up is the
basic structure of a randomized clinical trial \citep{byar_etal}. However, in situations where such
trials are not possible, one is often faced with {\it observational data} where the treatment
and outcome variables are measured without any randomization. A significant portion of the
field of causal inference is dedicated to assessing causality in observational data.

Causal inference has been in relatively common use in the climate sciences; however, application in 
meteorology has been more limited. Moreover, in both fields, the focus has been more on situations involving
time series data \citep{hannart_etal, hirt, kretschmer, massmann, nowack, sugihara, tsonis}, 
mainly because assessing causality in a temporal setting is relatively straightforward \citep{granger}.

One of the reasons why it is difficult to establish causality in non-temporal observational data is
the possibility that the relationship between the treatment and the outcome involves a third variable,
often referred to as a lurking variable; see Appendix A of the accompanying article
\citep{marzban_match} for a glossary of common terms in causal inference. If the treatment causes the 
lurking variable, which in turn causes the outcome,
the lurking variable is said to mediate the relationship between the treatment and the outcome.
If both the treatment and the outcome are caused by the lurking variable, then one says that the
treatment-outcome relationship is confounded (by the lurking variable), and the lurking variable
itself is called a confounder. The present article deals with the estimation of the causal effect
in the presence of confounding.

Given that the methods of causal inference have been developed across a wide range of fields, a simple
taxonomy is not readily available. However, a few distinguishing characteristics can be identified;
for example, some methods assume the confounders of interest are fully measured, while others address
situations in which known confounders are not measured or only partially observed.  A method in the
former category, called matching, is demonstrated in \citep{marzban_match}; the present work considers a method for the latter
situation, specifically one that is based on the notion of an Instrumental Variable (IV)
\citep{angrist_imbens_rubin}.\footnote{Angrist and Imbens were awarded a portion of the 2021 Nobel
Prize in Economics, for their methodological contributions to the analysis of causal relationships.}
Although there exist many methods for addressing unmeasured
confounders, the IV method is selected for demonstration here because, at its core, it is based
on regression, a common method in meteorology circles.

A well-known example of the utility of the IV method can be found in
\citep{richardson_tchetgen} where the goal of the study is to assess the causal effect
on life span of the radiation accompanying the atomic bombs dropped on Hiroshima and Nagasaki.
That study shows that when
numerous measured confounders are included in a regression model, the effect of the radiation
from the bombs on life span is not statistically significant at the 0.05 level. However, that
conclusion is reversed when an IV method is employed to account for unmeasured confounders.

IV methods have also been employed in climate sciences, especially with regard to the effect
of changes in the climate on quantities of social interest. For example, \citet{jiang}
use IVs to assess the impact of fish production on piracy. \citet{huang} apply
IV methods to assess the causal impact of heat and air pollution on mortality. \citet{wu_etal}
use an IV to study the effect of typhoons on labor markets.  Another example, with less social
focus, is by \citet{sun_etal}, who study the relationship between cloud cover and surface air
temperature. In addition to their focus on climate and social metrics, another important
distinguishing characteristic of these works, in contrast to that presented here, is that all of these
studies involve time series data. 

The outline of the paper is as follows: After a brief introduction to the history and taxonomy of
causal inference, the method section begins by reviewing one framework - the potential outcomes
framework.  Given the prominence of conditional expected values in that framework, the next
subsection examines the conditions under which multiple regression can be used to estimate the
average treatment effect. One such condition is whether or not the confounders are measured.
The sister article accompanying this article considers the former case through a
method called matching; the present article demonstrates the IV method, designed to allow for
the estimation of the average treatment effect even if confounders are not measured, when certain
conditions are met.
The Data section describes the process by which the data are selected not only to align the
two articles, but also to allow for a comparison of multiple estimates of the causal effect.
The Results section begins by using a simulated data set to demonstrate the equality of the
estimates under ideal conditions and proceeds to compare the various estimates based on
real data. The paper ends with a summary of the conclusions, a discussion
of further details and assumptions, and a set of proposals for future work. The aim of this
and the sister article is to encourage the meteorology community to examine methods of causal inference,
in non-time-series settings, for a better understanding of the causal structure underlying
meteorological variables in the presence of confounding. The pedagogical nature of both papers
implies that neither work provides a complete introduction to the field, and that the 
cited books and papers ought to be consulted for more detail.

\section{Method}

\subsection{The Theory}

It is important to begin by acknowledging that the causal inference methods described here
and in the sister article do not address the question of whether a treatment variable $A$
has a causal effect on an outcome variable $Y$. Indeed, the causal
relationships between variables must be established {\it a priori}, based on theoretical
considerations and domain knowledge. For example, see Figure 1, where the direction of the arrows
displays the direction of the causal relationships. Only after the causal structure of the
variables has been defined, one can apply the methods of causal inference to estimate and test the causal
effect, based on data. Moreover, some authors even advise against using methods of causal inference
if it is not possible, at least in principle, to envision performing the study as a randomized 
experiment \citep{cochran, dorn, holland, rubin_a, rubin_b, rubin1974, rubin2007}.

\begin{figure}[t]   
\center
\includegraphics[height=1.5in,width=5in]{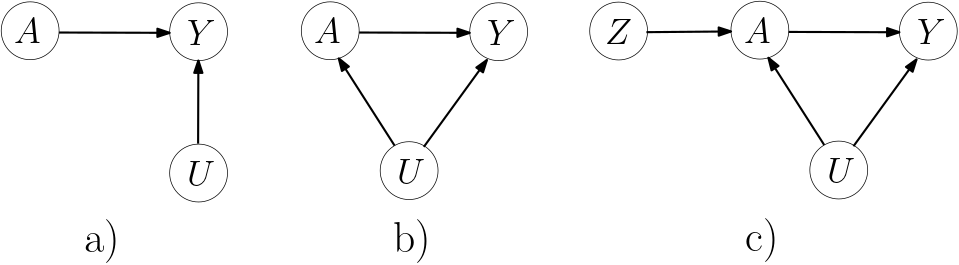}
\caption{Diagrams displaying situations where (a) $Y$ is affected by $A$ and an ``error term'' $U$, 
(b) $U$ is a common cause of both $A$ and $Y$, and (c) there exists an instrument $Z$.}
\end{figure}

A great deal of the work in the field of causal inference is dedicated to determining the conditions
required for causality to be identified.  If the theoretically-justified causal structure is one
resembling that displayed in Figure 1a, where a treatment $A$ causally affects the outcome $Y$,
subject to some unmeasured ``error'' $U$, then the assessment of causality is relatively
straightforward. However, in the presence of a confounding variable $U$, affecting both the
treatment and the outcome (Figure 1b), certain conditions must be met.\footnote{In the sister
article, the confounders are denoted $X$; here, they are denoted $U$ because they are assumed to be
unmeasured.} To address some of these
conditions, it is convenient to consider the case of a binary treatment, denoted $A=0$ and $A=1$,
even though the bulk of the present work deals with a continuous treatment.  According to
the potential outcomes framework, one measure of causal effect is the
Average Treatment Effect (ATE), defined as $ATE = E[Y(1)] - E[Y(0)]$, where $Y(0)$ and $Y(1)$ are the
potential outcomes if the experimental unit is exposed to the two treatment values, respectively.

The first condition necessary for causal inference is that the conditional expectation
of each potential outcome, given the corresponding treatment, is equal to the conditional expected
value of the observed outcome, all for a specific value of the confounder $U$. Mathematically, this
assumption is written as
\begin{equation}
E[Y(A) | A, U] = E[Y | A, U]\;,
\end{equation}
and is a consequence of the consistency assumption described in the sister article. The second
necessary assumption
is that, conditional on $U$, the treatment and the potential outcomes are independent; mathematically,
\begin{equation}
(Y(0), Y(1)) \perp A \; | \; U\;.
\end{equation}
Under this condition, for a given value of $U$, the treatment is essentially randomized across the
subjects, and so confounding can be removed by conditioning on $U$. This condition, often stated as
the No Unmeasured Confounding Assumption (NUCA), and also called unconfoundedness or the ignorability
assumption, effectively means that the covariate vector $U$ appearing in Eq. (2) includes all
confounders. A violation of this assumption implies confounding, and therefore, leads to a biased
estimate of ATE\@.

The fundamental problem in causal inference is that one of the
two potential outcomes $(Y(0), Y(1))$ is unobserved.  The first assumption provides a direct link
between the unobserved potential outcomes and the observed outcome. The second condition aims to
eliminate confounding effects by breaking the dependence of the potential outcomes on the treatment.
If these assumptions are satisfied, then a method referred to as g-computation, or the g-formula,
can be employed to show that \citep{robins1986}
\begin{equation}
ATE = \sum_U \; (E[Y | A=1, U] - E[Y | A=0, U]) \; P(U) \;,
\end{equation}
where $P(U)$ denotes the distribution of the confounder $U$. In other words, $ATE$ can be estimated
by considering the observed difference in means for a given value of the confounder, and then 
averaging that difference across the values of the confounder. Given that these assumptions
involve the potential outcomes, they cannot be tested by data; their validity can be assessed only
from domain knowledge \citep{ding_book}.\footnote{The reader is encouraged to read the Concluding
Remarks in the sister article, where several additional assumptions are discussed.}

In this subsection the treatment variable $A$ has been assumed to be binary only for simplicity
of presentation. Henceforth, however, the treatment variable is allowed to
be continuous, in which case the ATE can be formulated as $E[Y(a')] - E[Y(a)]$ for some fixed
$a,a'$, or as the so-called incremental causal effect $E[Y(A+\delta)] - E[Y(A)]$
\citep{kennedy, rothenhausler}, or the derivative effect curve
$a \mapsto \frac{d}{da} E[Y(a)]$ \citep{zhang_chen}. Indeed, it is the derivative effect curve
that serves as the most natural generalization of ATE in the context of continuous treatments. The
estimation of ATE is immensely simplified in a linear constant-effect model setting, because then 
the regression coefficient of $A$ measures a constant marginal causal effect. In turn, the
regression coefficient measures $ATE$ if it is interpreted as the linear constant effect per 
one-unit increase in the treatment variable; see the Concluding Remarks section in this paper for
further discussion.

\subsection{Multiple Regression}

Given the appearance of conditional expectations, it is natural to model them with regression.
Reconsider the simplest example (Fig. 1a). If $U$ is unmeasured, then one may write
$Y = A \beta + U$. For generality, $A$ may be
an $n \times p$ matrix of $n$ observations on $p$ treatment variables, in which case $\beta$ denotes
a $p \times 1$ column-vector of parameters. Under the assumption that the ``error'' $U$ is uncorrelated
with $A$, Eq. (3) implies  $ATE = \beta$, which can be estimated by the
Ordinary Least Square (OLS) estimate
\begin{equation}
\hat{\beta}_{OLS} = (A^T A)^{-1} A^T Y.
\end{equation}

It is useful to point out that although the OLS estimate is often derived by minimizing the
sum of squared errors, that criterion is equivalent to the condition $A^T \hat{U} = 0$,
where $\hat{U} = Y - A \hat{\beta}$. This view of regression, where the estimation criterion is
$A^T \hat{U} = 0$, is convenient because it readily leads to the OLS estimate via simple matrix
multiplication. For example, left-multiplying both sides of $Y = A \hat{\beta} + \hat{U}$ by $A^T$,
and using $A^T \hat{U} = 0$, leads to the OLS estimate in Eq. (4).
Here, this estimate of ATE will be referred to as the OLS estimate. If $A$ were binary, the
OLS estimate would be equal to the difference in means used in the sister article.

In the hypothetical case when the variable $U$ is measured, one may
write $E[Y|A, U] = A \beta_a + U \beta_u$, or equivalently, $Y = A \beta_a + U \beta_u + \epsilon$,
where $A, Y$, and $U$ are $n \times 1$ column-vectors of $n$ observations (assumed to be standardized,
for convenience), and $\epsilon$ denotes the ``error term.'' In this case, Eq. (3) implies
$ATE = \beta_a$.  In other words, the conditions that allow for the assessment of causality are
sufficient for allowing the regression coefficient to be interpreted as a measure of causal effect.
$\beta_a$ is then a measure of ATE after adjusting for the measured confounder.  Defining the 
projection operator $Q = I - U (U^T U)^{-1}U^T$, where $I$ denotes the identity matrix, 
the adjusted OLS estimate of ATE is
given by
\begin{equation}
\hat{\beta}_{OLS+adj} = (A^T Q A)^{-1} A^T Q Y.
\end{equation}

A desirable property of an estimator is statistical consistency (unrelated to the aforementioned
assumption with the same name), i.e., $\hat{\beta} \rightarrow \beta$, as $n \rightarrow \infty$.
It is straightforward to show $\hat{\beta}_{OLS} = \beta + (A^TA)^{-1}A^TU$.
It is evident that $(A^TA)^{-1}A^TU = (\frac{1}{n} A^TA)^{-1}(\frac{1}{n}A^TU)$, which by the law of
large numbers converges to $E[(A^TA)^{-1}] E[A^TU]$, which in turn is zero if $E[A^TU]=0$. In short,
the OLS estimator of $\beta$ is statistically consistent, if $A$ and $U$ are uncorrelated. 
Similarly, it can be shown that the OLS+adj estimator is also statistically consistent under a 
similar condition.

However, if $U$ represents variables that are in fact correlated with (or have a causal effect on)
the treatment $A$, then $E[A^TU]=0$ cannot be assured, and so the OLS estimator is statistically
inconsistent. Figure 1b depicts such a situation. By similar reasoning, the OLS+adj estimator is
statistically inconsistent if the error term represents confounders that are correlated with the
treatment.  In short, in the presence of confounding, the OLS and OLS+adj estimators of ATE are biased
and statistically inconsistent. Under relatively mild conditions (e.g., if the variance of the
estimator is bounded), statistical consistency implies asymptotic unbiasedness (i.e., unbiasedness
for large samples), and for that reason, the more serious defect of the OLS and OLS+adj estimators is
their statistical inconsistency.  A solution to this problem is provided next.

\subsection{Instrumental Variable}

The IV method assumes the existence of a ``fourth'' variable, often called an instrument, denoted $Z$ 
in Fig. 1c. Here, all of the variables are assumed to be continuous; \citet{swanson_etal} develop
the IV methodology for situations where all of the variables are binary. A good instrument
1) causes, and is therefore correlated with the treatment $A$, 2) is independent of, and hence 
uncorrelated with, the (unmeasured) confounder $U$, and 3) does
not have a direct effect on the outcome variable $Y$, or equivalently, the effect of $Z$ on $Y$
is only ``through'' $A$. These three conditions are often referred to as relevance, independence, and
exclusion restriction, respectively.

To get a sense of the mathematical relevance of these assumptions, consider the following derivation 
of a commonly used estimator of the ATE in the IV setting:
Left-multiplying $Y = A \beta + U$ by $Z^T$ yields $Z^T Y = Z^T A \beta + Z^T U$.
The assumption that $Z$ is uncorrelated with $U$ implies that $Z^T U = 0$ (in expectation), and the
assumption that $Z$ is correlated with $A$ implies that $Z^T A$ can be inverted, leading to
\begin{equation}
\hat{\beta}_{IV} = (Z^T A)^{-1} Z^T Y = \frac{Cov[Z,Y]}{Cov[Z,A]},
\end{equation}
where the last expression is known as the Wald estimator.

It is straightforward to show
$\hat{\beta}_{IV} = \beta + (Z^T A)^{-1} Z^TU$, and a similar argument to the OLS case above implies
that $\hat{\beta}_{IV}$ is a statistically consistent estimator of $\beta$ if $E[Z^TU]=0$.  As such,
the ATE can be
estimated consistently, even in the presence of unmeasured confounders $U$, if one can identify and
measure a variable $Z$ that satisfies $E[Z^TU]=0$, i.e., an instrument that is uncorrelated
with the confounders. The equivalence of ATE and the parameter estimated in Eq. (6)
is contingent on several other assumptions that are discussed in the Concluding Remarks section.

Technically, the second assumption is not testable, because in the IV setting the confounder $U$ is
not measurable. The third assumption is also not testable, because in its precise, mathematical
statement it involves the potential outcomes one of which is inherently unobservable.  Indeed, note
that the derivation leading to Eq. (6) does not require the third assumption at all. Practically, the
third assumption is necessary because it assures that the estimated effect of $A$ on $Y$ is not 
further ``confounded'' by any effect that $Z$ may have on $Y$.  

The above ``trick'' for constructing a consistent estimator works if $(Z^T A)$ is invertible, which
requires it to be a square matrix, implying that $Z$ and $A$ have the same matrix
dimensions. In other words, the number of the treatment variables and instruments, as well as the
number of observations of each, must be the same. This condition is often referred to as the
``just identified'' case. A solution to this restriction is provided by the method of 
Two-Stage-Least-Squares (TSLS), described in Appendix A, according to which a statistically
consistent estimate of ATE is given by Eq. (A5).\footnote{Note that in the just identified case, 
and with the instruments normalized such that $Z^TZ=1$, the expression in Eq. (A5) reduces to that 
in Eq. (6).}
As such, not only can the number of instruments be larger than the number of treatment variables,
but more importantly, the data sets from which $A$ and $Y$ are obtained may be entirely different
from the data set containing the observations on $A$ and $Z$ \citep{angrist_krueger}. This feature
of TSLS regression can be beneficial in meteorology where data on different variables
may be in different archives. 

To summarize, the ``magic'' of the IV method is in allowing the estimation of ATE in the presence of 
unmeasured confounding variables. The only requirement is assurance from theory and/or domain
knowledge that the IV satisfies 
the conditions stated above. Technically, two additional conditions must also be assumed; they are 
often addressed separately because they do not pertain to how ATE is estimated, but rather to how it 
is interpreted.  
Finally, it is also important to point out that the IV assumptions allow for assessing causality, often 
called ``identification of a causal effect'' in causal inference circles. But the IV assumptions do 
not imply that a regression coefficient equals the ATE\@. That equivalence additionally relies on a 
linear structural model and the assumption that the causal effect is constant across the units.
See the Conclusions and Discussion section for further elaboration of these subtle issues.

\section{Data}

The present paper is one of two articles intended to demonstrate two different causal inference 
methods - matching and the IV method. To align the two papers, both employ the same data set. 
Details of the
data are described in the sister article \citep{marzban_match}. Suffice it to say that the data are
monthly averages from the North American Regional Reanalysis (NARR) archive \citep{mesinger},
consisting of 418 variables on a $149 \times 117$ grid for January 2001. The only difference between
the two data sets, necessitated by the assumptions underlying the two methods, is that the matching 
method described in the first paper requires the treatment variable to be binary; the IV method allows 
for a continuous treatment variable, and therefore, the treatment variable in the present paper is no 
longer converted to a binary variable.

The variables used in this study are as follows:
\begin{itemize}
\item Treatment variable $(A)$ = downward shortwave radiation flux $(W/m^2)$, referred to as
``radiation,'' when appropriate.
\item Outcome variable $(Y)$ = surface potential temperature $(^\circ C)$,  referred to as ``temperature,'' when appropriate.
\item Instrumental variables $(Z)$ = cloud cover (\%), at three levels: low, medium, and high.\footnote{In NARR, the three 
   levels are defined as follows in terms of atmospheric pressure $p$:\\
   Low: $p > 642 mb$ (roughly from the surface to $3.5 km$);
   Medium: $642 \geq p \geq 350 mb$ (roughly 3.5 to $8 km$);
   High: $p < 350 mb$ (roughly above $8km$).}
\item Confounding variables $(U)$ = geopotential height $(gpm)$ at 150 and $875 hPa$, referred to as ``height,'' when appropriate.
\end{itemize}
Theoretical justification as well as histograms and scatterplots for the treatment, outcome, and 
confounding variables are presented in the sister article. The IV method does not involve measured 
confounders, but the availability of measured confounders, necessary for the matching method, 
allows one to compute the confounder-adjusted estimate of ATE, denoted OLS+adj above. This estimate provides a value with which the IV estimate can be compared.
In short, the use of measured confounders in this article is due 
only to the pedagogical nature of the two complementary articles.

To set the stage for application to the NARR data, a simulated data set is also examined.

\subsection{Theoretical Justification}

Recall from Figure 1c that a perfect IV ($Z$) must satisfy (at least) three conditions: It must 1) 
cause, and therefore be correlated with, the treatment $(A)$; 2) be uncorrelated with the 
confounder $(U)$, and 3) have an effect on the outcome $(Y)$ only through the treatment. Only the first 
condition can be verified empirically, and therefore, theoretical justification is necessary for the
choice of instrument; see the Conclusion and Discussion section for more detail.

Regarding the first condition, cloud cover represents an obvious candidate for an instrument 
acting on the treatment (downward solar radiation) due to its role in blocking and ultimately 
reducing the solar radiation flux at the surface.  However, the following considerations suggest 
that of the three levels of cloud cover the mid-level cloud cover may be more appropriate as an 
instrument. 

As shown in the sister article \citep{marzban_match}, the analysis domain extends from about 
$30^\circ$ to $60^\circ N$. The wintertime storm track tends to reside in the central latitudes of 
this range, leading to the expectation that the middle-tropospheric (mid-level) cloud cover associated 
with extra-tropical cyclones should generally be greater at those latitudes. Near the southern edge 
of the analysis domain there tend to be lesser mid-level cloud cover fractions due to the sinking 
branch of the Ferrel cell in the sub-tropics. Across the higher latitudes of the analysis domain the 
downward solar radiation is small, and so the presence versus absence of clouds makes little 
difference in the net surface heat flux in an absolute sense. This implies that the longitudinal 
variations in the mid-level cloud cover in the central range of latitudes should be particularly 
important in modulating the downward solar radiation. 

It bears noting that the low- and high-level cloud coverages are not liable to be as suitable as
instruments. The low-level cloud coverage is a poor candidate because of its direct linkages to surface
potential temperatures. In some situations it leads to warming by increasing the downward longwave
radiation and in other situations cooling, often in association with precipitation and not just
radiative effects. Moreover, it is expected to be more highly correlated (in a negative sense) with
geopotential heights, especially at pressure levels in the lower troposphere (e.g., $875 hPa$).
The high-level cloud coverage is also not a good choice as an instrument largely because it has
considerably weaker impact on the downward solar radiation at the surface relative to the much
more optically thick cloud decks at the medium and lower levels.  

Regarding the second IV condition, consider that extra-tropical cyclones tend to be much more 
prevalent upstream of ridges of higher geopotential height in the middle and upper troposphere,
with a reflection at the 150 hPa pressure level, where 
upward motion is favored, and less common and in earlier stages of their development immediately 
downstream of these ridges where downward motion commonly occurs. It is known that the mean wintertime 
flow over North America features a ridge of higher geopotential heights from the middle troposphere 
into the lower stratosphere over western North America due to the combined effects of the large-scale 
terrain, namely the Rocky Mountains, and land versus ocean effects. Therefore, for similar values of 
the $150 hPa$ geopotential height in the central latitudes of the analysis domain, it can be posited 
that there should be greater mid-level cloudiness to the west of the ridge axis over the coastal 
portion of 
western North America, and lesser cloudiness over and immediately downstream of the Rocky Mountains, 
with some recovery (moderate cloud cover values) downstream over eastern North America. The shift in 
the pattern of the mid-level clouds relative to the $150 hPa$ geopotential height distribution 
should result in a minimal spatial correlation between the two fields.

As for the third condition, consider that mid-level clouds are well above the lower troposphere 
and therefore cannot directly influence surface potential temperatures. They are associated
with processes that can have remote effects on surface temperatures. For example, mid-level
clouds often include ice crystals that are important for the initiation of particles large
enough to grow into droplets or ice crystals large enough to fall as precipitation,
with this precipitation then having the potential to have diabatic effects near the surface,
generally in the form of cooling. Mid-level clouds also typically result in increased
emission of downward longwave radiation from their bases in the middle troposphere than
otherwise in their absence. The importance of this enhancement in the downward
longwave radiation to the net heat fluxes at the surface and ultimately surface temperatures
depends strongly on the humidity and fraction of cloud cover in the lower troposphere.
Generally speaking, for the analysis domain used here and during the month of January,
mid-level clouds occur much more frequently than precipitation, and therefore their
impacts on surface temperatures in the mean, aside from their role in reducing the downward solar
radiation, are apt to be manifested through their influence on the downward longwave radiation.
As stated above, the latter influence is not necessarily substantial, and hence the mid-level
cloud cover represents a reasonable choice for an instrument with surface potential temperature
as the outcome variable. 

\subsection{Statistical Considerations}

Among the three main IV conditions, the first is testable, at least in terms of correlation. Stated 
more accurately, the correlation between the IV and the treatment can be used to assess the strength
of the association between them.  To that end, Table 1 shows the 95\% confidence interval for the 
correlation between the treatment (radiation) and the three IVs (low-, mid-, and high-level cloud 
cover). When this correlation is low (high), one says that the instrument is weak (strong); it can be 
shown that a weak instrument can lead to a biased estimate of ATE; see \citep{crown_bias} and the 
Conclusion and Discussion section here. According to Table 1, IV3 is the weakest of the three IVs, 
with negligible association with the treatment, consistent with the theoretical expectations stated
above. 

Based on the correlations in Table 1, IV1 is the strongest of the three, with IV2 somewhat weaker than 
IV1, both with significant associations with the treatment. As such, it may seem that IV1 should lead 
to the least biased estimates of ATE\@. However, that statistically based conclusion ignores not
only the theoretical considerations against IV1 as a good instrument, but also the other two IV 
conditions, both of which are untestable. In other words, a good IV can be justified only based on 
theoretical considerations.

However, in this pedagogical setting where data on the confounders are available, one can compute 
the correlation between the IVs and the confounders; recall that a good IV must have low correlation 
with the confounders. According to Table 1, the IV least correlated with both confounders is IV2,
again consistent with the theoretical considerations. 

Although theoretical and statistical considerations point to IV2 as the better of the three instruments,
here all three IVs are examined for pedagogical reasons - specifically, to
demonstrate the effect of poor IVs on the estimated ATE\@.

%
%

\begin{table}[t]
\caption{The 95\% confidence interval for the correlation between some of the variables.}
\begin{tabular}{c|ccc}    
          & IV1 (low-cloud) & IV2 (mid-cloud) & IV3 (high-cloud) \\ \hline
treatment (radiation) & (-0.49, -0.47)  & (-0.42, -0.40)  & (-0.04, -0.01) \\
confounder 1 (height at 150 hPa)  & (-0.46, -0.44) & (-0.20, -0.18) & (0.31, 0.33)  \\
confounder 2 (height at 875 hPa)  & (-0.54, -0.52) & (-0.18, -0.15) & (0.30, 0.33) \\
\end{tabular}
\end{table}

Relevant histograms and scatterplots are shown in Figure 2 only for IV2 (mid-level cloud cover)
and the confounder at $150 hPa$. It can be seen that the treatment (radiation) is correlated with 
the outcome (temperature) with correlation coefficient $0.75\; (95\% \; CI: \; (0.745, 0.758))$, as 
required for causal inference. The remainder of the scatterplots confirm the correlations discussed 
in Table 1. The non-random structure in the scatterplots is a direct consequence of the spatial 
structure of the fields; the maps for the treatment, outcome,
and confounding variables are shown in the sister article, and the maps for the three IVs are
shown in Figure 3.

\begin{figure}[t] 
\center
 \includegraphics[height=5in,width=5in]{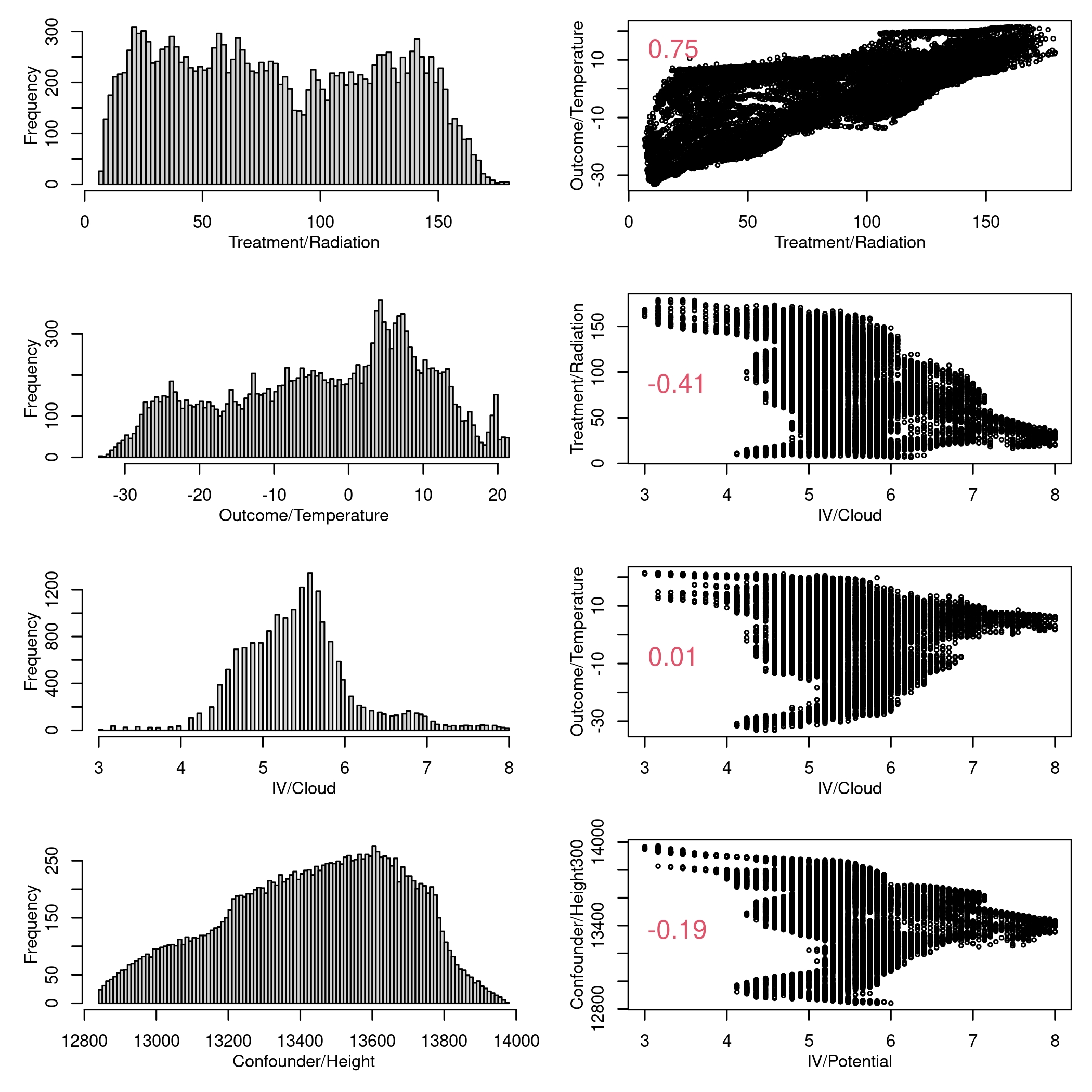}
\caption{Left: Histograms of the treatment variable (radiation), the outcome variable (temperature),
IV2 (mid-level cloud cover), and one confounder (height at $150 hPa$). Right: pairwise scatterplots 
for some of the variables, and the corresponding correlation coefficient (in red).}
\end{figure}

\begin{figure}[t]   
\center
 \includegraphics[height=3.5in,width=8in]{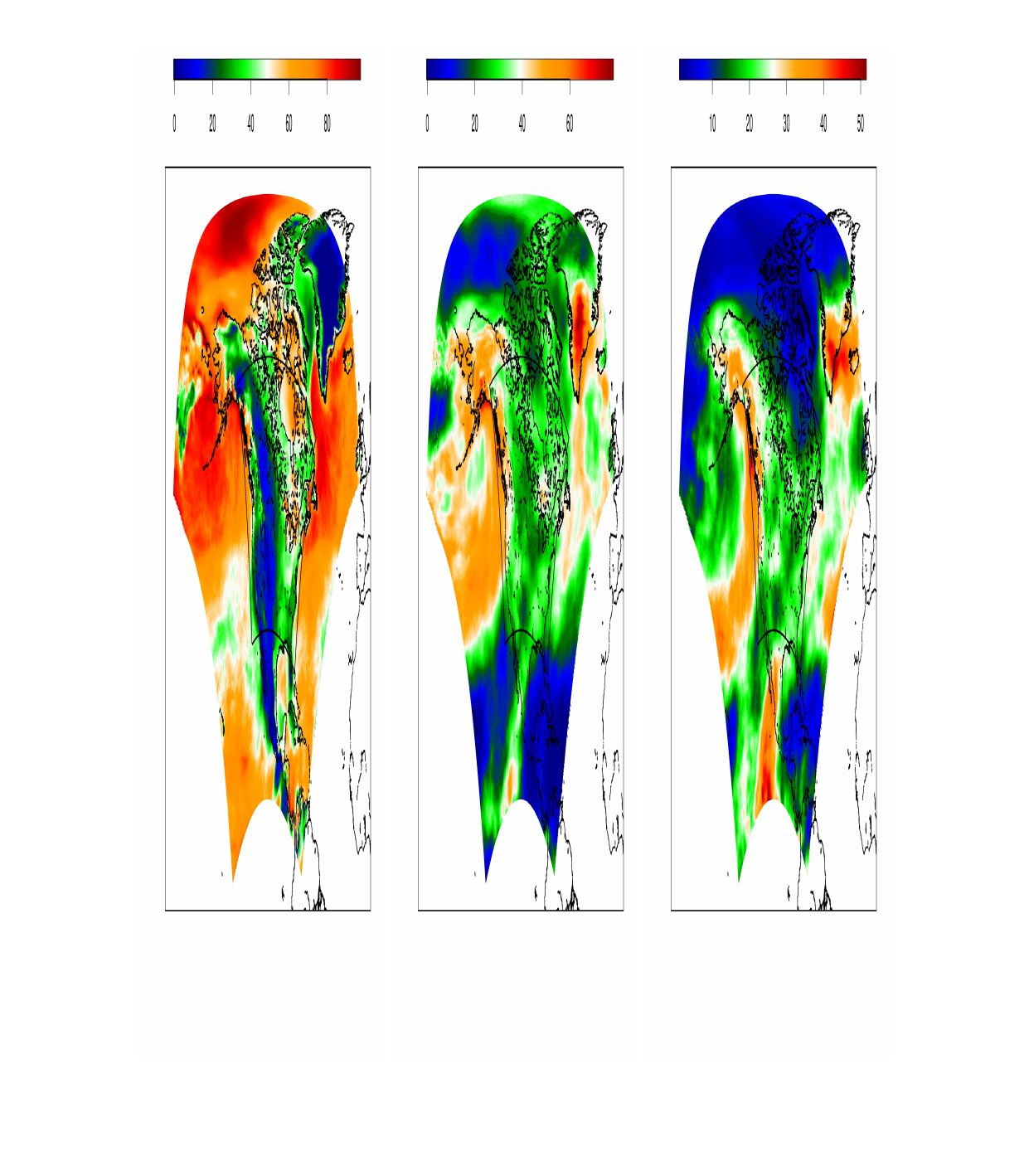}
\caption{From left to right, the map of the three IVs, low-, medium-, and high-level cloud cover (\%) 
for January 2001.  The inner domain shows the region of analysis.}
\end{figure}

\section{Results}

To summarize, three estimates of ATE are computed. The simplest is the OLS estimate (Eq. (4)),
based on the assumption that there are no confounders. In the presence of confounders, this estimate is 
inconsistent. Assuming the confounders are
measured, then another estimate is provided by the OLS+adj estimate in Eq. (5). A third estimate is
provided by the IV estimate in Eq. (A5), without use of the confounders. 
These estimates are referred to, respectively, as the OLS, the OLS+adj, and the IV estimates.  
Again, in the absence of any confounding, all estimates will be equal.
And if all confounders are known and measured, then the OLS+adj and the IV estimates will be equal.

The analysis is performed in R, using the ivreg package \citep{Rcore}. The main function
for developing an IV model is ivreg(). Given that all ATE estimates can be written
in closed form, analytic confidence intervals are also computed for all ATE estimates.
R code for analyzing the simulated and NARR data are provided in the supplementary material
for this paper.

\subsection{Results from Simulated Data}

For a simple demonstration, 1,000 samples of size 100 are generated on variables $A, Y, Z$, and $U$ 
(Fig. 1c), satisfying 
\begin{eqnarray}
E[Y|A,U] &=& A + 2 U \;, \\
E[A|Z,U] &=& 2 Z + 3 U \;,
\end{eqnarray}
and with the following correlation matrix:
\begin{equation}
\left( \begin{array}{ccccc}
  &   A  &  Y   &  U   & Z    \\
A & 1.00 & 0.87 & 0.64 & 0.43 \\
Y & 0.87 & 1.00 & 0.73 & 0.29 \\
U & 0.64 & 0.73 & 1.00 & 0.00 \\
Z & 0.43 & 0.29 & 0.00 & 1.00
        \end{array} \right).
\end{equation}
Note that as seen in Eq. (7), $ATE = 1$.
The supplementary material for this paper provides R code for both generating and modeling this
simulated data.
First, note that in this simulated data set the confounder $U$ is measured.
Also, note that the elements of the correlation matrix are consistent with the general IV setting:
The treatment and the outcome are highly correlated $(0.87)$, and $U$ is strongly correlated with both
the treatment $(0.64)$ and the outcome $(0.73)$, consistent with $U$ being a confounder.
$Z$ is highly correlated with the treatment $(0.43)$, and weakly correlated
with the confounder $(0.00)$. As such, the three IV assumptions are reasonably satisfied.

\begin{figure}[t] 
\center
\includegraphics[height=3in,width=3in]{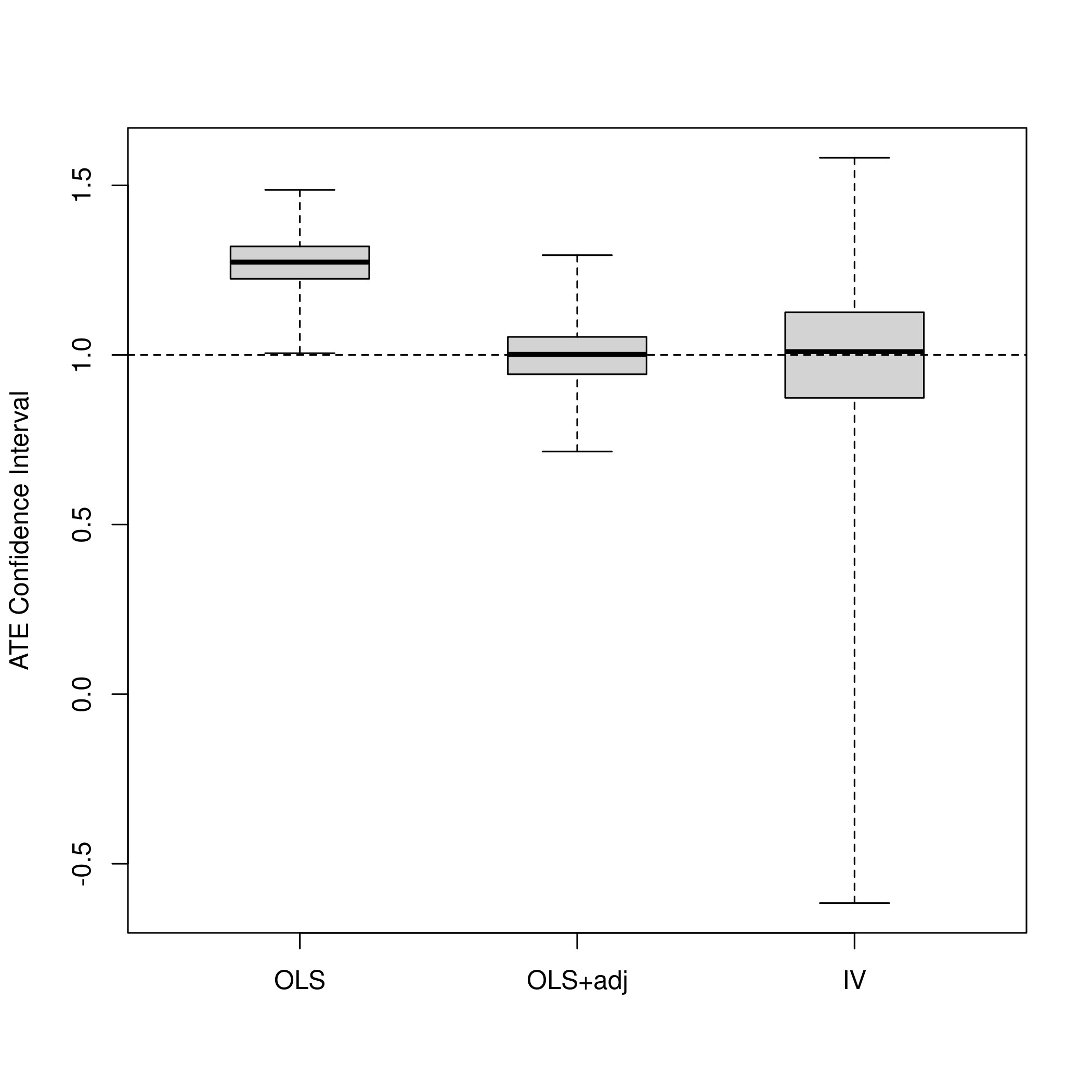}
\caption{Boxplots of simulated ATE values according to the OLS estimate, the OLS+adj estimate, and
the IV estimate. The horizontal line corresponds to the true value $ATE=1$.}
\end{figure}

Boxplots of the 1,000 ATE values, from each of the three estimates, are shown in Figure 4.
Based on the OLS boxplot, OLS over-estimates the true ATE of 1.0, consistent with the fact that the
presence of the confounder $U$ leads to biased and inconsistent estimates of ATE\@. The remaining
estimates have boxplots that are centered on the true value of ATE\@. This centering is a consequence
of the idealized nature of the simulated data and is not guaranteed in real data. It can
also be seen that the IV estimates have the largest variability of the three estimates. This loss
of precision is a consequence of using an estimate that does not require measured confounders.

\subsection{Results from Real Data}

The 95\% confidence intervals for ATE, based on the three estimates, are shown in Figure 5. 
The labels on the x-axis indicate which IV or IVs are used in the IV estimate. The OLS estimate (in black)
does not depend on the choice of the IV, and is therefore constant across the figure.
Also constant is the OLS+adj estimate, which adjusts for the two confounders (geopotential height
at $150 hPa$ and $875 hPa$). This estimate is consistently lower and with the opposite sign compared to the 
OLS estimate. This reversal in sign is addressed in the sister article as an instance of Simpson's 
paradox. 

As for the IV estimates (in blue), the results vary greatly depending on which IV or in what
combination they are used. Indeed, the estimate based on IV3 (High\_Cloud) are so extreme that the
values displayed in Figure 5 are ATE/30. The large magnitude for this estimate of ATE 
is a direct consequence of the near-zero denominator in the Wald estimate (Eq. (6)), and is
consistent with the theoretical considerations in section 3.1. Theoretical considerations also weigh
against IV1 (low-level cloud cover), which is seen to lead to an ATE estimate
even larger than the OLS estimate. In short, both of the IVs that violate the IV conditions lead to
anomalous estimates of ATE\@. 

As such, only the ATE estimate from IV2 is considered reliable, and it can be seen to be consistent with 
that in the sister article. That comparison is necessarily qualitative because
the IV estimates here are based on a continuous treatment variable, while those in the sister article
are derived from a binary version of the same treatment variable.

As an attempt to assess how the estimate of ATE is affected by a combination of good and bad
instruments, the last column in Figure 5 (denoted ``Low+Mid'') shows the ATE estimate if IV1 and IV2
are used jointly. As seen, the estimate remains inconsistent with the expected ATE\@. Again, 
this result emphasizes the importance of selecting good instruments; see the Conclusion and Discussion
section for more detail.

A meteorological explanation for the results in Figure 5 is as follows: High-level clouds
mostly affect longwave (downward) radiation through emission (the greenhouse effect).
Such clouds are not particularly efficient at blocking incoming shortwave radiation, so high
clouds have a net warming effect. In contrast, low-level clouds are substantially more effective at
attenuating shortwave radiation by reflecting solar radiation back to space (albedo). Hence they
result in lower daytime temperature which dominates the daily temperature cycle. Because longwave
radiation is a direct function of surface temperature, there is less upward longwave
radiation with low clouds present. Mid-level clouds are important for both reflection of shortwave
radiation back to space and emit longwave radiation downward towards the surface simultaneously,
so the correlations with the confounders (Table 1) are small for mid-level clouds, 
which helps to satisfy the IV conditions.

\begin{figure}[t] 
\center
\includegraphics[height=3.5in,width=3.5in]{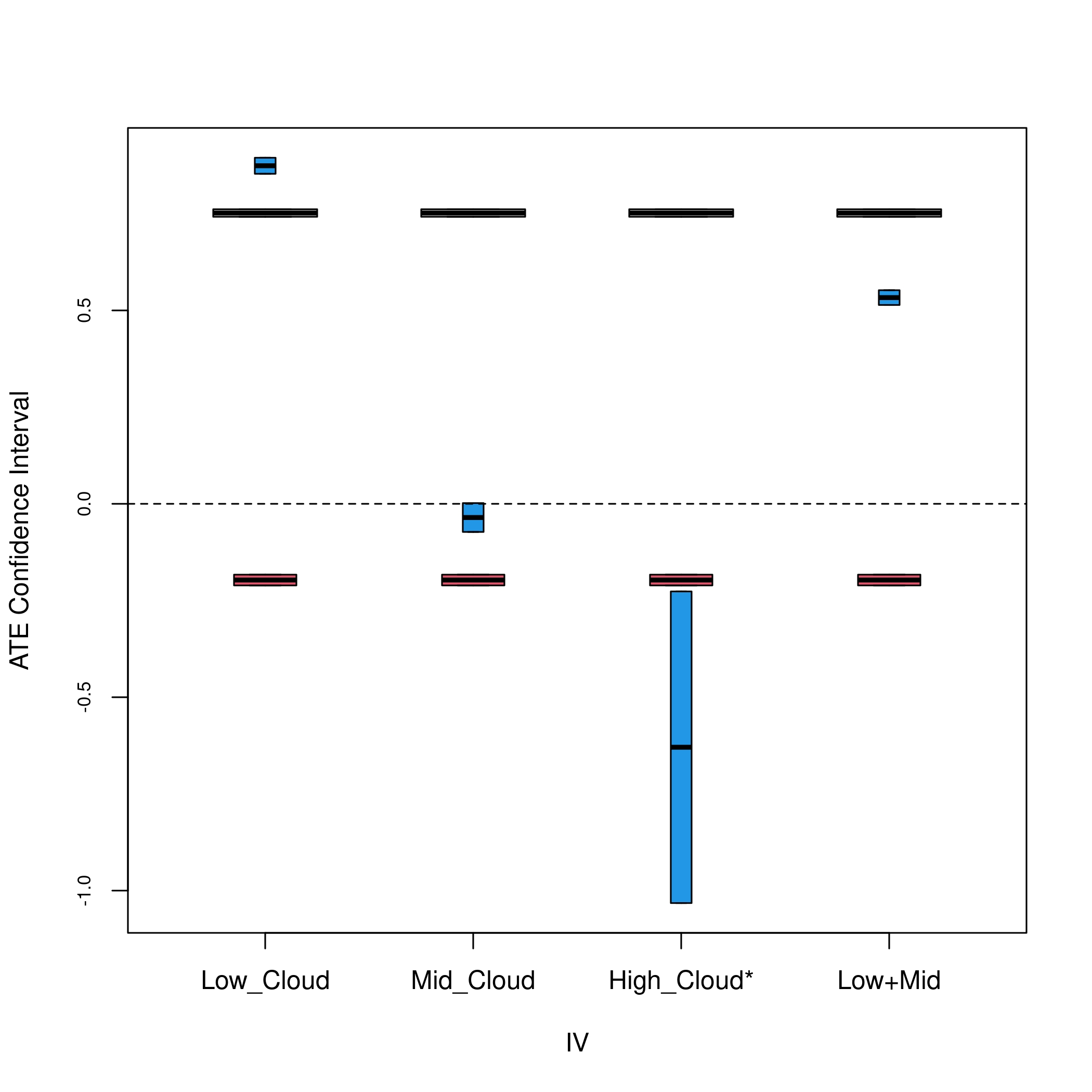}
\caption{The 95\% confidence interval for ATE according to the OLS estimate (in black), the 
OLS+adj estimate (in red), and the IV estimate (in blue). The first three columns indicate the
specific IV used (i.e., low-, mid-, and high-level cloud cover), and the last column shows the ATE 
estimate if low- and mid-level cloud cover are used simultaneously. 
*The CI for high-level cloud cover is divided by 30.}
\end{figure}

\conclusions[Conclusions and Discussion]

In this second of two articles on regression-based causal inference, the method of Instrumental
Variables (IVs) is demonstrated on gridded, meteorological data. Both papers aim to assess the causal 
effect of downward shortwave radiation flux on potential temperature, after the confounding effect of 
geopotential height has been taken into account. Whereas the matching method demonstrated in the sister
article assumes the confounding variables are measured, the IV method assumes the existence of 
confounding variables that are not measured. Here, qualitative agreement is found between the
two papers, but only after close attention is paid to theoretical justification for a good IV.
Unlike in simulated data where the IV method is shown to work as expected, on real data there is 
large variability across good and bad IVs in terms of the estimated causal effect.
This finding is not surprising given that \\
1) IV estimates are not based on any observations on the confounders; \\
2) as stated at the outset, theoretical considerations must be taken into account to justify
the validity of an IV; and\\
3) the IV method described here is the simplest among the family of IV methods designed to
provide more reliable estimates of ATE\@.\\
These points are further addressed below. Armed with the methods described in the two papers,
and R code provided in the supplementary material, meteorologists can quantify the causal
structures underlying atmospheric phenomena.

Many of the substantive extensions that can be made to the present work are the same as those listed
in the sister article. Briefly, it is important to account for spatial and/or temporal structures in 
data. Spatial confounding is considered by \citet{dupont1, dupont2, gilbert}, and the IV method in the 
presence of spatial structure is considered in \citep{reich_etal}. Additionally, it is useful to 
partition the analysis based on topography and/or seasons, and to consider other confounders and 
instruments.

The following subsections provide guidance on how to deal with the difficulties in finding good 
instruments.

\subsection{Strong and Valid IVs}

As mentioned in the IV section of the paper, the first IV assumption (Relevance) can be tested
empirically by examining the correlation between the instrument and the treatment. An instrument 
with a low correlation with the treatment is said to be weak. An instrument that fails to satisfy 
the other two assumptions (Independence and Exclusion Restriction) is said to be invalid.\footnote{Some
authors define an invalid instrument as one that fails all three IV assumptions, 
e.g., \citep{baiocchi}.} Either weak or invalid instruments lead to biased estimates of ATE; see 
Section 16.5 in \citep{hernan_robins}.

The study of weak instruments is an active area of research. For example, Section 23.6.3 in 
\citep{ding_book} discusses an alternative to TSLS which is more robust when
dealing with weak instruments, though at the cost of being more computationally intensive.
A statistical test has also been proposed, based on the F-distribution, for testing weak 
IVs \citep{windmeijer_weakIV}. \citet{stock2000gmm} develop the asymptotic 
theory for the generalized method of moments, and propose a confidence interval for ATE that accounts 
for weak instruments. 

Assessing the validity of instruments is also an active area of research. Section 6.1 of the IV 
tutorial by \citet{baiocchi} provides several methods for assessing the extent to which one or
both of the validity assumptions are satisfied. Section 6.2 of \citet{baiocchi} also discusses numerous 
methods for sensitivity analysis in the IV setting, including two specific tests, one for the 
independence assumption and another for the exclusion restriction assumption. \citet{kiviet} argues 
that, although the IV assumptions cannot be fully tested, there are certain situations in which one 
can gain some useful information on the validity of the assumptions. Further work on sensitivity 
analysis can be found in \citep{cinelli_hazlett1, cinelli_hazlett2, oganisian}.

\subsection{Searching for Valid IVs}

Most methods for selecting valid IVs fall into one of two categories: in one class, the problems
deal with a single instrument, while in the other there exists a set of potential instruments. 
As an example of the former, \citep{pearl2009causality} shows that the IV assumptions lead to so-called
instrumental inequalities that can be used to falsify an instrument; these turn out to be
necessary conditions, but not sufficient. When dealing with multiple instruments, it can be shown
that by considering a slightly different notion of independence, one can arrive at generalized 
instrumental inequalities that test the independence assumption; these inequalities turn out
to be both necessary and sufficient \citep{brito, kedagni}.

\citet{kang2016} develop an algorithm for estimating causal 
effects if more than half of the instruments are known to be valid, and without knowing which half.
By introducing other structures on a set of instruments, it is even possible to develop data-based
methods for selecting valid IVs \citep{singh_iv_search_ml, guo_iv_search}. The manner in which
this line of research avoids the fact that some IV assumptions are not testable is by imposing
structure on the IVs, and then examining the consequences of the imposed constraints.

Finally, to emphasize that simple, correlation-based tests of IV validity do not exist, note that
it would be wrong to assess the exclusion restriction assumption by testing the partial 
correlation between the IV and the outcome, given the treatment; in fact, given the structure of the 
graph in Figure 1c, that correlation is not zero because the treatment variable appears as what is 
called a collider. It can be shown that conditioning on a collider induces a correlation between 
the variables that cause it (here, the instrument and the confounders), and hence between the 
instrument and the outcome, even if the instrument is valid; see, e.g., the last two paragraphs in
Section 7.2 in \citep{morgan_book}. Similarly, it would be wrong to test whether the IV and the 
outcome are uncorrelated, because such a zero correlation would lead to a zero value for the Wald 
estimator in Eq. (6). In Section 19.2.3, \citet{huntington_effect} discusses several
attempts at ``testing'' IV validity assumptions; the following quote from that book adequately
summarizes how the tests should be done:

``A more reasonable use of these tests is when you have an instrument that you are really certain 
about, but are worried that maybe you've just had a bad luck of the draw. Using these tests to 
disappoint yourself about an instrument you were pretty certain on is a much more justifiable use 
of them than using them to reassure yourself about an instrument you're uncertain on. 
Good ol' statistics, always looking for a way to disappoint.''

\subsection{Concluding Remarks}

In IV literature, in addition to the basic assumptions of causal inference (stated in the sister
article), and the three IV assumptions mentioned previously, two other assumptions are often discussed, 
called homogeneity and monotonicity.
The former, in its simplest form, requires that the effect of the treatment $A$ on the outcome $Y$ is
constant across all units. A statement of the second condition requires considering whether or not
a unit (e.g., a person) will adhere to the treatment to which that unit is assigned. This condition is
automatically satisfied in the current application because a grid point does not have the option
of not adhering to the treatment (radiation) it receives. Regardless, both conditions address the 
question of whether the estimated causal effect pertains to the entire population (i.e., all units/grid 
points) or to a specific subpopulation. For example, when homogeneity does not
hold, but monotonicity does, the TSLS estimate of ATE is specific to a subset of the
population called compliers; see Sections 16.3 and 16.4 in \citep{hernan_robins}.

It is also important to clarify the causal parameter estimated by the IV estimator in
Eq. (6). In general, the Wald or TSLS estimator does not identify the usual population-level ATE
(e.g., as considered in the sister article). With a binary instrument, and under the standard IV 
assumptions together with the aforementioned monotonicity assumption, the IV estimand is commonly 
interpreted as a local average treatment effect (LATE), i.e., the average causal effect among units 
whose treatment status is affected by the instrument. With a continuous instrument, as in the
present article, the analogous object is more naturally described through a local IV curve, which 
characterizes causal effects for subpopulations whose treatment level is shifted
locally by the instrument \citep{kennedy_robust}. Thus, without further assumptions, the IV 
estimand should be understood as a local causal effect rather than the usual ATE\@.
In the present paper, however, the linear structural model is assumed to be
$A = Z\alpha + V, \; Y = A\beta + U,$ under which $\beta$ represents a constant causal effect of
$A$ on $Y$. Under this homogeneity assumption, the distinction between local and population-level
causal effects disappears, and the IV estimand can be interpreted as the average causal effect per
one-unit increase in $A$. As such, throughout this work ``ATE'' is shorthand for this
constant causal effect in the IV setting.\footnote{Returning to the monotonicity assumption, rather 
than interpreting it literally in terms of human compliance, in this physical meteorology setting, 
one may argue that the assumption is satisfied through this local effect interpretation of ATE\@.}
Nonlinear extentions of the IV method are examined in \citep{abadie_semiparametric}.

It is known that the IV estimator can have finite-sample bias but is asymptotically consistent
\citep{ding_book}. The extent of the bias
depends on a variety of factors, but the quality of the instrument is one of the more influential
factors. As a result, a great deal of the IV literature has been dedicated to the study of
weak instruments \citep{andrews, keane}. Additionally, TSLS has been compared
with other competing methods in terms of the reduction of bias resulting from unmeasured confounders
\citep{zhang_lewsey}.

We emphasize that the present description of the IV method represents only an example of how it may
be employed in meteorological applications. In particular, we chose treatment and outcome variables
whose relationships are plausible from a meteorological perspective, but otherwise unremarkable.
Different combinations of variables may be explored, and we hope that the present work will motivate
such efforts. The substantive findings may vary with the atmospheric circulation; different outcomes
are apt to be associated with periods of strong climate phenomena (e.g., a sample featuring a strong
ENSO state), for other times of year, using time series rather than a single month, and changes to
the spatial domain that may affect how the findings generalize.


\codeavailability{R Code for the analysis performed here is available as supplementary material.} 

\dataavailability{The NARR data are publicly available at the following URL:\\
https://www.ncei.noaa.gov/products/weather-climate-models/north-american-regional.} 




\appendix
\section{Two-Stage Least Squares}    

This appendix presents the details of the Two-Stage-Least-Squares (TSLS) method 
\citep{theil, basmann, angrist_imbens}.

Consider the structural equations describing Figure 1c.
Let $A$ denote the treatment, $Y$ the outcome, and $Z$ the IV\@. Then,
\begin{eqnarray}
Y &=& A \beta + U  \\
A &=& Z \alpha + V \; ,
\end{eqnarray}
where $U$ represents one or more unmeasured confounders, and $V$ is the error term for $A$.
Deviating slightly from Fig. 1c, for generality, $U$ and $V$ are allowed to be different. However,
if/when any of the variables in $V$ include (or are correlated with) the variables in $U$, then
one says that confounding exists.

Note that it is not always possible to estimate $\beta$ in (A1) via Ordinary Least Squares (OLS),
by using the IV condition that $Z$ is independent of $U$. Writing $Y = A \hat{\beta} + \hat{U}$,
along with the OLS criterion $Z^T \hat{U} = 0$, leads to $Z^T Y = Z^T A \hat{\beta} + Z^T \hat{U}$.
If $A$ and $Z$ have different matrix dimensions, then $Z^TA$ is not a square matrix and is therefore
non-invertible. The method of TSLS is designed to circumvent this obstacle and to estimate $\beta$.

As in OLS, in the first stage of TSLS the parameters in (A2) are estimated from
\begin{equation}
A = Z \hat{\alpha} + \hat{V} \hspace{1.0cm} s.t. \hspace{1.0cm}  Z^T \hat{V} = 0 \;,
\end{equation}
i.e., $\hat{\alpha} = (Z^TZ)^{-1} Z^TA$, and then the predictions of $A$ are found:
$\hat{A} = Z \hat{\alpha} = HA$, where $H=Z(Z^TZ)^{-1}Z^T$ is called the ``hat matrix.''

The second stage of TSLS develops another regression model, this time involving $Y$ and the
predictions $\hat{A}$:
\begin{equation}
Y = \hat{A} \hat{\gamma} + \hat{\epsilon} \hspace{1.0cm} s.t. \hspace{1.0cm}  \hat{A}^T \hat{\epsilon} = 0.
\end{equation}
It follows that
\begin{equation}
\hat{\gamma} = (\hat{A}^T \hat{A})^{-1} \hat{A}^T Y = (A^THA)^{-1} A^THY \;.
\end{equation}

It is now easy to show that $\hat{\gamma}$ in Eq. (A5) is a consistent estimator of $\beta$, i.e., the
ATE of $A$ on $Y$: using Eq. (A1) in Eq. (A5) gives
\begin{equation}
\hat{\gamma} \; = \; (A^THA)^{-1} A^TH (A\beta + U) \; = \; \beta + (A^THA)^{-1} A^THU \; = \; \beta + (A^THA)^{-1} A^TZ(Z^T Z)^{-1} Z^T U \;.
\end{equation}
Under the IV moment condition $E[Z^T U]=0$ and the rank condition that $E[Z^T A]$ has full column 
rank, the second term in (A6) converges in probability to zero, and so, the TSLS estimator is 
consistent for $\beta$.  As such, $\hat{\gamma}$ is a consistent estimator of $\beta$,
i.e., $\hat{\gamma} \rightarrow \beta$ when the sample size tends to infinity (see Section 23.5 in
\citet{ding_book}). Note that the
matrices whose inverse is required in TSLS are all square matrices, and therefore, usually invertible.
The $\hat{\gamma}$ in Eq. (A5) is the quantity used to estimate ATE in the IV setting.



\noappendix       





\appendixfigures  

\appendixtables   



\authorcontribution{Marzban and Zhang have contributed to the causal inference aspects of the work, and Bond and Richman have provided meteorological expertise.} 

\competinginterests{The authors declare that they have no conflict of interest.} 

\disclaimer{No funding has been provided for this work.} 

\begin{acknowledgements}
Zhenman Yuan is acknowledged for an exchange regarding TSLS.
\end{acknowledgements}




 \bibliographystyle{copernicus}
 \bibliography{paper.bib}

\end{document}